\documentclass[aps,prd,reprint,nofootinbib,superscriptaddress,preprintnumbers,amsmath,amssymb,latexsym,array,enumerate,letterpaper,twocolumn]{revtex4-1}

\usepackage{inputenc,amsmath,amssymb}
\usepackage{graphicx}
\usepackage{natbib}
\usepackage{xcolor}
\usepackage[colorlinks=true,urlcolor=brown,linkcolor=blue,citecolor=magenta]{hyperref}
\usepackage{graphicx}
\usepackage{comment}

\newcommand{\lsim}{{\;\raise0.3ex\hbox{$<$\kern-0.75em\raise-1.1ex\hbox{$\sim$}}\;}}
\newcommand{\gsim}{{\;\raise0.3ex\hbox{$>$\kern-0.75em\raise-1.1ex\hbox{$\sim$}}\;}}

\def\bea{\begin{eqnarray}}
\def\eea{\end{eqnarray}}
\def\beq{\begin{equation}}
\def\eeq{\end{equation}}
\def\beqnn#1\eeq{\begin{align*}#1\end{align*}}
\def\ba{\begin{array}}
\def\ea{\end{array}}
\def\bc{\begin{center}}
\def\ec{\end{center}}

\newcommand{\GeV}{\mathrm{\;GeV}}

\newcommand{\keV}{\mathrm{\;keV}}
\newcommand{\eV}{\mathrm{\;eV}}

\newcommand{\mpl}{M_P}
\newcommand{\Neff}{N_\textrm{eff}}

\newcommand{\Fig}[1]{Fig.~\ref{#1}}

\newcommand{\Eq}[1]{Eq.~(\ref{#1})}

\begin{document}

\title{Dynamical Generation of the Baryon Asymmetry from a Scale Hierarchy}

\author{Jae Hyeok Chang}
\email{jhchang@fnal.gov}
\affiliation{Theory Division, Fermi National Accelerator Laboratory, Batavia, IL 60510, USA}
\affiliation{Department of Physics, University of Illinois Chicago, Chicago, IL 60607, USA}
	
\author{Kwang Sik Jeong}
\email{ksjeong@pusan.ac.kr}
\affiliation{Department of Physics, Pusan National University, Busan 46241, South Korea}
	
\author{Chang Hyeon Lee}
\email{chlee@o.cnu.ac.kr}
\affiliation{Department of Physics and Institute of Quantum Systems, \\
Chungnam National University, Daejeon 34134, South Korea}	
\author{Chang Sub Shin}
\email{csshin@cnu.ac.kr}
\affiliation{Department of Physics and Institute of Quantum Systems, \\
Chungnam National University, Daejeon 34134, South Korea}
\affiliation{Center for Theoretical Physics of the Universe, \\
Institute for Basic Science, Daejeon 34126, South Korea}
\affiliation{Korea Institute for Advanced Study, Seoul 02455, South Korea}

\preprint{FERMILAB-PUB-24-0012-T-V}
	
\begin{abstract}
We propose a novel baryogenesis scenario where the baryon asymmetry originates directly from a hierarchy between two fundamental mass scales: the electroweak scale $v$ and the Planck scale $M_P$, in the form of 
\begin{equation}
Y_B \sim  \sqrt{\frac{v}{M_P}}  \, .  \nonumber
\end{equation}
This relation straightforwardly gives the observed baryon yield today $Y_B$, which can be a hint for underlying fundamental physics. We provide an example of baryogenesis models that yield this relation. Our model is based on the neutrino-portal Affleck-Dine mechanism, which generates the asymmetry of the Affleck-Dine sector during the radiation-dominated era and subsequently transfers it to the baryon number before the electroweak phase transition.
The observed baryon asymmetry is then a natural outcome of this scenario.
The model is testable as it predicts the existence of a Majoron with a keV mass 
and an electroweak scale decay constant. The impact of the relic Majoron on the effective number of neutrinos ($\Delta N_{\rm eff}$) can be measured through near-future cosmic microwave background observations.
\end{abstract}
	
\maketitle
	
\section{Introduction}
The observed baryon asymmetry \cite{Planck:2018vyg} is often parametrized with
\bea 
Y_B \equiv \frac{\bar{n}_B}{s} = (0.82 - 0.92) \times 10^{-10} \, ,
\eea 
where $\bar{n}_B$ is the net baryon number density and $s$ is the entropy density.
This value is significantly larger than the baryon asymmetry naturally anticipated in the Standard Model (SM), requiring new physics beyond the SM. 
However, if the baryon asymmetry arises from new physics, it is essential to realize the observed baryon asymmetry with the model parameters of the new physics.
Previous works have typically relied on small couplings or large wash-out effects to yield the observed baryon asymmetry \cite{Bodeker:2020ghk,Elor:2022hpa,Barrow:2022gsu}. 
	
In this article, we propose a scenario of baryogenesis where the baryon asymmetry results directly from a scale hierarchy between the two important mass scales in phenomenology: the electroweak scale 
$v= 246$~GeV and the reduced Planck mass $\mpl = 2.4 \times 10^{18} \GeV$, in the form of
\bea 
\label{eq:YB-v-Mpl}
Y_B = {\cal O}(0.01) \sqrt{\frac{v}{\mpl}} \sim 10^{-10}\, .
\eea 

A relation between an observable and a scale hierarchy may not be a simple coincidence but a hint for underlying fundamental physics. As a known example, the number of hydrogen atoms in the Sun $N_{H,\odot}$ is given by the scale hierarchy between the Planck mass $M_P$ and the mass of hydrogen $m_H$: $N_{H,\odot} \sim M_\odot/m_H  \sim (M_P/{m_{\rm H}})^3$.
This relation is not accidental but results from a balance between the gravitational and degenerate pressure known as the Chandrasekhar limit. If the number of atoms were much larger than this number, the Sun would have been unstable from gravitational collapse, falling into a black hole. Thus, a typical long-lived star such as the Sun has a similar mass to this limit. In this manner, we regard the coincidence that the baryon asymmetry is expressed with the two well-known scales as a hint for new physics and give an example of working models for this scenario.

We utilize the Affleck-Dine (AD) mechanism \cite{Affleck:1984fy,Dine:1995kz}, where the asymmetry is generated from the dynamics of a $B-L$ charged complex scalar field $\phi$. $\phi$ naturally has an electroweak scale mass, $m_\phi = {\cal O}(100\GeV)$, if the scalar potential of $\phi$ is governed by the same mechanism that ensures the stability of the Higgs boson mass against quantum corrections. 
In contrast to previous works on AD baryogenesis \cite{Allahverdi:2012ju}, we explore the generation of $B-L$ asymmetry during
the radiation-dominated era
and its subsequent transfer to the SM sector through neutrino-portal interactions. Therefore in our scenario, there is no dependence on the reheating temperature. 

A simple model realizing our scenario is constructed by 
employing supersymmetry (SUSY) \cite{susy1:KaneShifman,Ellis:1983ew,Nilles:1982dy}. Because SUSY ensures the cancellation of all orders of quadratic divergences in perturbation theory, the scalar potential can be well-organized without concern for various quantum corrections from unknown UV physics. The supersymmetric Lagrangian can be systematically derived from the superpotential and K\"ahler potential using the framework of superspace and superfields.~\cite{Salam:1974yz}.\footnote{
Spontaneous SUSY breaking in a hidden sector generates soft SUSY breaking terms, resulting in the superpartners of the SM particles being heavier than the electroweak scale.} 
The relevant interactions of the model in terms of superpotential are given by 
\beq\label{eq:model}
	W= y_\nu \ell h N + \frac{1}{2} \lambda_N \phi N^2  +      \frac{\kappa}{4M_P} \phi^4 + \cdots  \, ,
\eeq
where $\ell$ is the SM lepton doublet, $h$ is the Higgs field, $N$ is a right-handed (RH) neutrino with $B-L=1$, and $\phi$ is a SM singlet scalar field with $B-L=-2$. $y_\nu$ is the neutrino Yukawa coupling that gives neutrino masses $m_\nu$ through the seesaw mechanism \cite{Minkowski:1977sc,Ramond:1979py,Gell-Mann:1979vob,Yanagida:1979as,Mohapatra:1980yp}, while $\lambda_N$ and $\kappa$ are $\mathcal{O}(1)$ coefficients.
A global $U(1)_{B-L}$ symmetry is preserved at the renormalizable level, allowing only the seesaw operators, but is explicitly broken by Planck-scale suppressed operators as generally expected due to quantum gravity effects \cite{Holman:1992us,Kamionkowski:1992mf,Barr:1992qq}. We highlight how SUSY breaking can provide a well-organized scalar potential in the presence of explicit $U(1)_{B-L}$ breaking terms and show how the two fundamental mass scales are combined to determine the baryon asymmetry. Moreover, we establish an interesting connection between the observable dark radiation and the baryon asymmetry with certain implications of the electroweak scale, emerging as a generic prediction.

\section{Summary of cosmological history}
Initially, the SM plasma dominates the energy density of the Universe, while the abundances of the novel particles $\phi$ and $N$ are negligible. 
When the temperature drops to $T_{\rm AD} \sim \sqrt{m_\phi  \mpl}$, the AD mechanism becomes active, generating the asymmetry of $\phi$. 
The temperature continues to decrease, leading to the thermalization of the AD sector ($\phi$ and $N$) with the SM plasma at the temperature $T_N$ via the neutrino Yukawa interactions. Once thermalized, the asymmetry is transmitted to the lepton sector through the neutrino portal, and the baryon asymmetry is induced via the weak sphaleron process and frozen after the electroweak phase transition. 
After the $B-L$ phase transition happens, all novel particles decay to the Majoron $J$, the pseudo-Nambu-Goldstone boson associated with the spontaneous $U(1)_{B-L}$ breaking. The Majoron, along with its decay products, contributes to the effective number of neutrinos ($\Delta \Neff$), providing a phenomenological signal for the model.

\section{Asymmetry generation}
With the field decomposition of $\phi= \frac{1}{\sqrt{2}} r_\phi e^{i\theta_\phi}$, the net charge density of the scalar field $\phi$ becomes
\beq \label{eq:nphi}
\bar{n}_\phi =  i (\dot\phi^*\phi - \phi^*\dot\phi)= r^2_\phi \dot\theta_\phi \, ,
\eeq
which can be interpreted as a rotating scalar field carrying angular momentum in the field space.
In the early Universe, a nonzero charge density can arise from the ``kick" along the $\theta_\phi$ direction induced by the $B-L$ breaking potential term for a large value of $r_\phi$. 
In addition to the supersymmetric contribution given by $|\partial W/\partial\phi|^2$, the scalar potential incorporates soft SUSY breaking terms \cite{Witten:1981nf,Dimopoulos:1981zb,Susskind:1982mw,Girardello:1981wz,Hall:1990ac},
\beq \label{eq:soft}
\Delta V_{\rm soft} = m_\phi^2 |\phi|^2 - \left(\alpha m_\phi W(\phi)+ h.c.\right),
\eeq
where the model-dependent constant $\alpha$ naturally assumes a value of ${\cal O}(0.1-1)$. 
Coupled to the SM fermions $\psi_a$ through the higher dimensional operator arising from the K\"ahler potential,  
\beq \label{eq:Hubble_induced}
\Delta {\cal L} =\kappa_a \frac{|\phi|^2 }{M_P^2} \bar\psi_a i \gamma^\mu D_\mu \psi_a, 
\eeq
the field $\phi$ acquires an additional mass. During the radiation-dominated era, the above interaction generates an effective mass squared of $\phi$ proportional to $H^2= \rho_{\rm tot}/(3M_P^2) = \pi^2 g_* T^4/(90 M_P^2)$. Here $H$ is the Hubble expansion rate, and $g_*={\cal O}(100)$ is relativistic degrees of freedom. 
The overall coefficient of this Hubble term $\kappa_H$ is typically the order of unity \cite{Kawasaki:2011zi}.
Including all these contributions,
the potential terms governing the evolution of $\phi$ in the early
universe are given by 
\bea \label{eq:phipotential}
&& \hskip -0.5cm  V = (m_\phi^2 - \hskip-0.05cm\kappa_{H} \hskip-0.05cm H^2)|\phi|^2 - \frac{ \kappa \alpha m_\phi}{4 \mpl} \left( \phi^4 + \phi^{* 4} \right) 
+ \frac{\kappa^2}{\mpl^2} |\phi|^6 \nonumber\\ 
&& \hskip -0.12cm =  \frac{1}{2} (m_\phi^2- \hskip-0.05cm\kappa_{H} \hskip-0.05cm H^2)r^2_\phi 
- \frac{ \kappa \alpha m_\phi}{8M_P} r^4_\phi\cos 4\theta_\phi + \frac{\kappa^2}{8M_P^2} r^6_\phi \, ,	\nonumber\\
\eea
in the field basis where $\alpha$ is real.
The terms involving other scalar fields
are irrelevant because they are fixed at the origin
at high temperatures.
Here we take $ m_\phi^2>0$ and $\kappa_H > 0$.
It is worth noting that SUSY protects the quadratic term from quantum corrections above the SUSY breaking scale by making the contributions from fermionic and bosonic loops cancel each other. Additionally, SUSY ensures that the $B-L$ breaking quartic term is highly suppressed by a factor of $m_\phi/\mpl$. 
A suppressed $m_\phi/M_P$
ensures that the scalar potential, as described by \Eq{eq:phipotential}, remains flat enough, allowing us to implement the Affleck-Dine baryogenesis. 
 
At high temperatures where $H\gg m_\phi$, 
the minimum of the potential is predominantly determined
by the Hubble-induced mass term. 
Having a mass similar to $H$, the radial field $r_\phi(x)$ resides at 
\bea \label{eq:rvev}
\langle r_\phi \rangle= \left( \frac{4\kappa_H}{3 \kappa^2} \right)^\frac{1}{4} \sqrt{H \mpl} \, .
\eea
On the other hand, the angular field $\langle r_\phi\rangle  \theta_\phi(x)$  has an effective mass smaller than $H$, 
resulting in its position in the field space being nearly frozen at an arbitrary value.
As the Universe expands and $H$ decreases, the sign of the quadratic term inverts, and the scalar potential is lifted subsequently.
Due to the rapid lifting compared to the Hubble time, there is an increase solely in the potential energy, with a negligible change in the radial field value. 
As the quadratic term approaches zero, the quartic potential term imparts a mass of $\mathcal{O}(H)$ to the angular field. 
Then, for a typical value of $\theta_\textrm{in}$, an initial misalignment angle of $\theta_\phi$, the angular field starts to roll toward the minimum. 
However, with the diminishing radial field value due to rolling, the potential barrier height along the radial direction also decreases, allowing the traversal of the angular field across the barriers.
The radial and angular motions can be described through rotation within the two-dimensional complex field space, characterized by a specific angular momentum.

The equation of motion for $Y_\phi \equiv \bar{n}_\phi/s$ is given by
\beq\label{eq:eom_nphi}
\frac{dY_\phi}{dt}  =- \frac{1}{s}\frac{\partial V}{\partial\theta_\phi} \, ,
\eeq
where the RHS of \Eq{eq:eom_nphi} serves as the origin of
the net $\phi$ charge density. Its impact is maximized just after the lifting of
the scalar potential at temperature  $T_{\rm AD}\sim \sqrt{m_\phi M_P}$ and gives $\dot\theta_\phi\sim \alpha m_\phi$ at $r_\phi \sim \sqrt{m_\phi M_P}$. 
After the onset of scalar field rotation, the $B-L$ breaking effect gets suppressed, freezing $Y_\phi$.

The value of $Y_\phi(t)$ can be obtained by integrating the equation of motion of $n_\phi$ in \Eq{eq:eom_nphi} over $t$ with the potential \Eq{eq:phipotential} and the initial condition $Y_{\phi}(t_{\rm in})=0$ in the radiation-dominated era:
\bea\label{eq:Yphitfin}
     Y_\phi (t) &=& - \int_{t_{\rm in}}^{t } \frac{1}{s} \frac{\partial V}{\partial \theta_\phi} dt'  \\
     &=& -    \left(\frac{\kappa \alpha m_\phi}{8 M_P}\right) \left(\frac{90}{\pi^2 g_*}\right)^\frac{1}{4}
     \int_{t_{\rm in}}^{t} \frac{ r_\phi^4(t') \sin{4 \theta_\phi(t')}  }{(H(t') M_P)^{3/2}} \, dt' \, .\nonumber\\ \label{eq:Yphitfinfull}
\eea
For the generation of asymmetry, the relevant dynamics arises when $H(t)=\frac{1}{2t} \sim m_\phi$, so we define $t_*$ with $ H(t_*)= m_\phi$ and rewrite \Eq{eq:Yphitfinfull} with a dimensionless time variable $\tilde t= t/t_*$,
\begin{widetext}
\beq  \label{eq:Yphiform}
   \tilde{Y}_\phi (\tilde t) \equiv Y_\phi(t=\tilde{t} \, t_*)  = - \frac{1}{6} \left( \frac{90}{\pi^2 g_*} \right)^{1/4} \left( \alpha\frac{\kappa_H}{\kappa} \sin{4 \theta_{\rm in}} \right) \sqrt{\frac{m_\phi }{M_P}}  
  \int_{\tilde{t}_\text{in}}^{\tilde{t}} \frac{\tilde{t}'^{3/2}}{2}
  \frac{\tilde{r}_\phi^4(\tilde {t}') \sin{4 \tilde{ \theta} _\phi(\tilde{ t}')}}{ r_*^4 \sin{4 \theta_{\rm in}}}\, d \tilde{t}' \, . 
\eeq
\end{widetext}
Here, $r_*= \langle r_\phi\rangle $ in Eq.~(\ref{eq:rvev}) with $H=m_\phi$, $\tilde{t}_{\rm in}=t_{\rm in}/t_*$, and $\tilde{r}_\phi(\tilde{t})$ and $\tilde{\theta}_\phi(\tilde{t})$ are $r_\phi(\tilde{t} \, t_*)$ and $\theta_\phi(\tilde{t} \, t_*)$, respectively.

One can perform a numerical calculation for the integration part, but we can start with analytic calculations with several approximations. The radial and angular fields show different characteristics before and after the rolling of the fields. For $\tilde t \ll 1$, the angular field is frozen, and the radial field follows Eq.~(\ref{eq:rvev}), which gives $r_\phi(t)\propto t^{-1/2}$, so we get $\tilde{r}_\phi^4(\tilde t) \sin 4\tilde{\theta}_\phi(\tilde t)/(r_*^4 \sin 4\theta_{\rm in})\approx \tilde t^{-2}$. 
On the other hand, for $\tilde t\gg 1$, the real and imaginary parts of the complex scalar field $\phi$ are oscillating as the independent harmonic oscillators with the angular frequency $m_\phi$. At the same time, the average value of $r_\phi ( t)$ decreases as $a^{-3/2}\propto  t^{-3/4}$, therefore $r_\phi(t) \propto t^{-3/4} \cos{m_\phi t}$.
The initial eccentricity is close to the unity so $\theta_\phi(t)$ remains $\theta_{\rm in}$ near the maximum of the oscillations. Thus we get $\tilde{r}_\phi^4(\tilde t) \sin 4 \tilde{\theta}_\phi(\tilde t)/(r_*^4 \sin 4\theta_{\rm in}) \simeq \tilde t^{-3} (\cos \frac{1}{2} (\tilde{t}-1))^4$ for $\tilde t\gg 1$. Then the final $Y_\phi$ at $t \rightarrow \infty$ can be separated and analyzed as
\begin{widetext}
\begin{eqnarray}\label{eq:Yphi}
  Y_\phi &\approx & -\frac{1}{6} \left( \frac{90}{\pi^2 g_*} \right)^{1/4} \left(\alpha\frac{\kappa_H}{\kappa}   \sin{4 \theta_\textrm{in}} \right)\sqrt{\frac{m_\phi }{ M_P}}  \left(  \int_{0}^{1}    \frac{1}{2 \tilde t^{1/2}}  \, d \tilde{t} + \int_{1}^{\infty} \frac{(\cos{\frac{1}{2}(\tilde t-1)})^4}{2\tilde t^{3/2}}   \, d \tilde{t}  \right) \nonumber \\
   &\approx& -\frac{1}{6} \left( \frac{90}{\pi^2 g_*} \right)^{1/4} \left(\alpha\frac{\kappa_H}{\kappa}   \sin{4 \theta_{\rm in}}\right) \sqrt{\frac{m_\phi }{ M_P}} \Big( 1+ 0.48 \Big)  \nonumber \\
   &\approx& -0.1 \, \alpha \frac{\kappa_H}{\kappa}  \sin{4 \theta_{\rm in}} \left(\frac{200}{g_*(T_{AD})}\right)^{1/4} \sqrt{\frac{m_\phi }{ M_P}}  \, ,
\end{eqnarray}
\end{widetext}
where the integration range of is simplified to $(0,\infty)$ because the initial and final time values are far from $\tilde{t}=1$. 

\begin{figure}[t]
    \centering
 {{\includegraphics[width=0.48\textwidth]{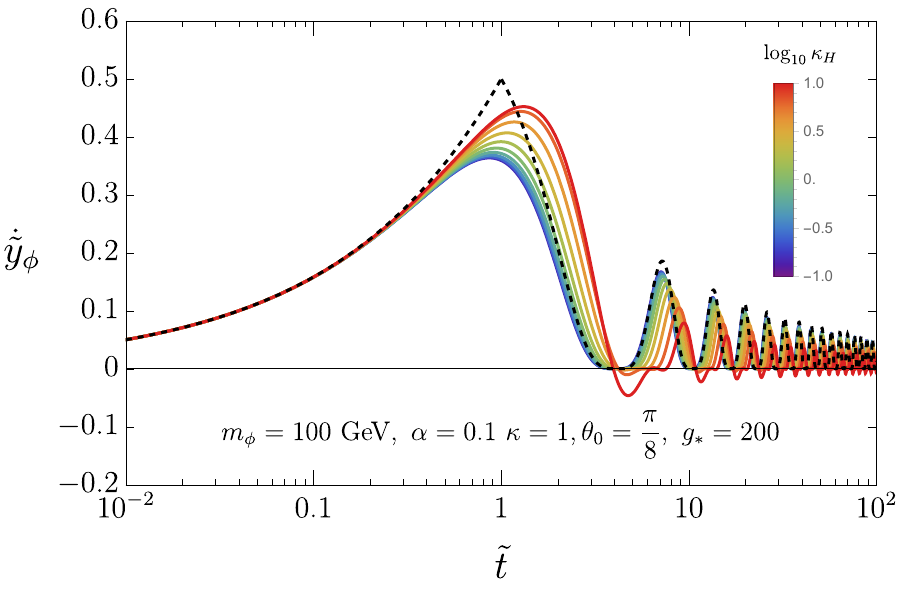}}}
    \caption{The $\kappa_H$-dependence of $\dot{\tilde{y}}_\phi(\tilde t)$  defined by \Eq{eq:smally} in radiation dominated era. 
    The black dashed line is the analytic approximation given by Eq.~(\ref{eq:Yphi}) that shows a good agreement with the actual value from the solution of $\phi(t)$ to Eq.~(\ref{eq:eomphi}) with a small deviation around $\tilde t=1$.     This is true for other values of $\kappa_H$.  
    }
    \label{fig:S1}
\end{figure}
\begin{figure*}
\centering
\hskip-0.3cm\includegraphics[width=0.7\textwidth]{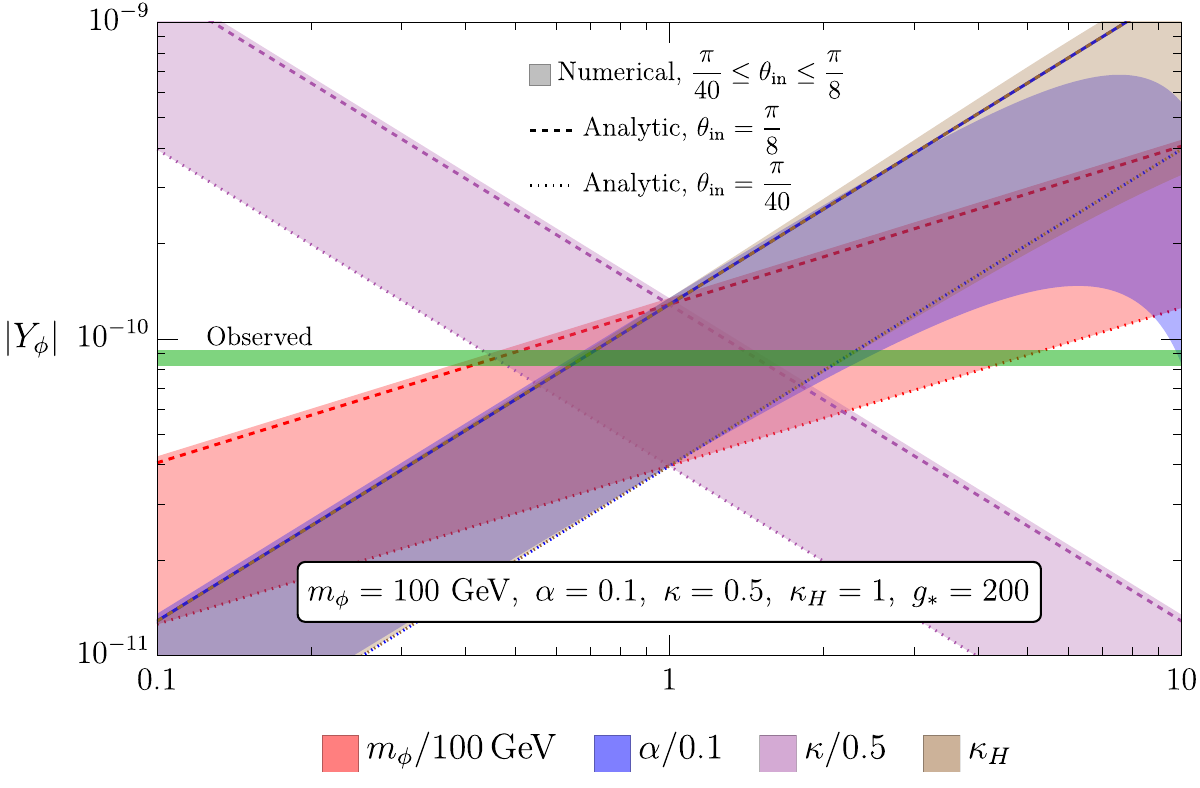} 
\caption{Analytic and numerical results for $|Y_\phi|$. The fiducial parameters are shown at the bottom of the plot, and the colored lines and bands show the case where one of the parameters is varied. We show analytic results in \Eq{eq:Yphi} for $\theta_\textrm{in}=\pi/8$ (dashed line) and $\theta_\textrm{in}=\pi/40$ (dotted line), and numerical results (bands) for $\theta_\textrm{in}$ values between them. $\theta_\textrm{in}=\pi/8$ corresponds to the maximum $Y_\phi$, while $\theta_\textrm{in}=\pi/40$ is the minimum within $10\%$ tuning. Note that the observed asymmetry value is shown in the green band and the labels of the horizontal axis for different colors are shown below the plot.}
\label{fig:Yphiplot}
\end{figure*}

The numerical value of $Y_\phi$ is obtained from the following process. From the \Eq{eq:phipotential}, the equation of $\phi(t)$ is
\begin{eqnarray} \label{eq:eomphi}
     \ddot{\phi}+3H \dot{\phi} +(m_\phi^2 -\kappa_H H^2) \phi -\frac{ \kappa\alpha m _\phi }{ M_P} \phi^{*3} + \frac{3 \kappa^2}{M_P^2} \phi^3 \phi^{*2}=0 \, . \nonumber\\
\end{eqnarray}
Then, we can evaluate \Eq{eq:Yphitfinfull} with
\begin{eqnarray}
     r_\phi^4 \sin{4 \theta_\phi} = -2i ( \phi^4 - \phi^{*4}) 
\end{eqnarray}
for the numerical solution of $Y_\phi(t)$.
\Fig{fig:S1} shows analytic and numerical evolutions of
\beq\label{eq:smally}
\dot{\tilde{y}}_\phi(\tilde t) \equiv 
\frac{d}{d \ln{ \tilde{t} } } \left( \frac{\tilde{Y}_\phi(\tilde t)}{\tilde{Y}_\phi(\tilde{t}=1)}\right)
\approx
\frac{\tilde{t}^{5/2}}{2}
\frac{\tilde{r}_\phi^4(\tilde t) \sin{4 \tilde{\theta}_\phi(\tilde t)}}{ r_*^4 \sin{4 \theta_{\rm in}}} \, .
\eeq
The black dashed line represents the analytic approximation given by \Eq{eq:Yphi} and colored real lines are numerical values with the solution of $\phi(t)$. Here, we note that the quadratic potential of $r_\phi(t)$ changes its sign at $t=t_{\rm flip}$, where $\kappa_H H^2(t_{\rm flip}) = m_\phi^2$. Depending on the size of $\kappa_H$, $t_{\rm flip}$ is different from $t_*$, but we confirm that $t_*$, rather than $t_{\rm flip}$, is the time when the behavior of $Y_\phi$ changes significantly, as shown in \Fig{fig:S1}.
We also display the final analytical and numerical results for $Y_\phi$ in \Fig{fig:Yphiplot}, which shows how well the parameter-dependence of $Y_\phi$ in the analytical expression agrees with the numerical results. 

The quantity $Y_\phi$ remains conserved throughout the rotational evolution of the scalar field.
Following the decay of $\phi$, the net $\phi$ number efficiently transfers to the actual baryon number in the visible sector.
For an electroweak scale $m_\phi$, a typical value of $\theta_\textrm{in}$, and the naturally expected range of model parameters 
$\alpha, \kappa_H, \kappa ={\cal O}(0.1-1)$, 
$Y_\phi$ attains ${\cal O}(10^{-10})$, which agrees with the experimentally observed $Y_B$.
We have checked that \Eq{eq:Yphi} is consistent with the numerical calculations for interesting parameter ranges, and the summarized results are depicted in \Fig{fig:Yphiplot}.

In Eq.~(\ref{eq:Yphi}), we see how the Sakharov conditions \cite{Sakharov:1967dj} are fulfilled.
The $B-L$ violating effect is encoded in $\alpha$, the coefficient of 
the $\phi^4$ potential.
The $C$ and $CP$ violations 
arise from the initial deviation of $\theta_\phi$
from the potential minimum.  
The asymmetry is quickly generated as $T$
crosses $T_{\rm AD}$ and becomes frozen after 
$\phi$ starts to oscillate, thereby satisfying the out-of-equilibrium condition.
These are the same as how the Sakharov conditions are satisfied in the conventional AD mechanism, however, 
Eq.~(\ref{eq:Yphi}) also shows a difference. In our scenario, the asymmetry does not depend on the reheating temperature since the AD mechanism happens during the radiation-dominated era. 

\section{Asymmetry transfer}  
The thermalization of $\phi$ with the SM thermal bath and
the subsequent transfer of asymmetry to the SM sector  
take place via the neutrino portal.
From Eq.~(\ref{eq:model}), 
the relevant interactions are given by 
\bea
\Delta {\cal L} = \bar N i\bar\sigma^\mu \partial_\mu N  - \left(y_\nu \ell h N + \frac{1}{2} \lambda_N \phi NN + h.c.\right) \, .
\eea 
Through the Yukawa coupling $y_\nu$, the RH neutrinos are produced from 
the SM thermal bath. The production of $N$ becomes cosmologically important when $\Gamma_N(T_N) = H(T_N)$, where
\bea
\Gamma_N \approx  4 \times 10^{-3} y_\nu^2 T \, 
\eea
is the production rate of $N$ from the SM plasma \cite{Besak:2012qm,Garbrecht:2013bia,Ghisoiu:2014ena}. 
The neutrino Yukawa coupling can be written as  $y_\nu = \sqrt{m_N m_\nu}/v$ for the RH neutrino mass $m_N$ in the vacuum.  Thus, $T_N$ is given by \cite{Escudero:2019gvw}
\bea 
T_{N} \sim 5\, m_N \left(\frac{\sum m_\nu}{0.05\,{\rm eV}}\right).
\eea 	
If $T_N > T_{\rm AD}$, the population of $N$ leads to the thermal potential of $\phi$ as $\sim \lambda_N^2 T^2 |\phi|^2$ that spoils the Affleck-Dine mechanism. However, we can easily control the thermal effect by taking a relatively small neutrino Yukawa coupling. 
For $m_N \ll \sqrt{m_\phi M_P}\sim 10^{\rm 10}\, {\rm GeV}$, we can safely ignore the direct effect of the amount of $N$ on the scalar potential of $\phi$. 
The indirect thermal effect exists when $\lambda_N|\phi| \gtrsim T$. From the renormalization group running of couplings, the mass of RH neutrinos affects the value of $y_\nu(T)$ at one-loop level, which leads to a logarithmic correction to the scalar potential \cite{Anisimov:2000wx} as $\Delta V(\phi) = {\cal O}(0.01) y_\nu^2 \lambda_N^2 T^4 \ln \lambda_N |\phi|/T$. 
For the scalar dynamics, it becomes insignificant when $y_\nu \lambda_N \lesssim 10^{-6} $. It falls within the parameter regime of interest to us.

During the thermalization of $N$ and $\phi$, we can neglect the slight change of the entropy density $s$ from new degrees of freedom in the thermal bath.
Before the onset of asymmetry generation,  the radial component of $\phi$
is positioned at the potential minimum given by \Eq{eq:rvev}.
During this phase, the energy density of $\phi$ is dominated by the homogeneous kinetic energy and is given by $\rho_\phi \sim H^3 \mpl$. 
Thus, it is negligible in comparison to the radiation energy density, $\rho_\textrm{rad} \sim H^2 \mpl^2$. 
After the asymmetry generation, $\rho_\phi$ scales as $a^{-4}$ shortly during free-rolling and as $a^{-3}$ after it starts to oscillate near the origin. 
Although $\rho_\phi$ decreases slower than $\rho_\textrm{rad} \propto a^{-4}$, $\rho_\phi$ remains negligible compared to $\rho_\textrm{rad}$ until they get thermalized. 
 
As the temperature becomes lower than $T_N$, $\phi$  and $N$ are all thermalized with the SM bath through the $\lambda_N \phi N N$ term, and asymmetry of $\phi$ is distributed to the SM lepton sector. 
The baryon asymmetry is also generated through the weak sphaleron process when the neutrino-portal interaction is in equilibrium before the electroweak phase transition ($T_{\rm EW}\approx 159\GeV$), more precisely before the freeze-out of the sphaleron process ($T_{\rm sp}\approx 132\GeV$) \cite{DOnofrio:2014rug}.  
This translates to the lower bound on the RH neutrino mass as $m_N\gtrsim 26\GeV$. The bound on $m_N=\lambda_N \langle \phi\rangle_{T=0}$ in the simplest model of neutrino implies that the neutrino mass is Majorana type, i.e. the $B-L$ number should be spontaneously broken in the vacuum. The spontaneous breaking of $U(1)_{B-L}$ in the early Universe should not lead to a wash-out of the existing asymmetry. 

After all AD sector particles decay, the asymmetry of the AD sector is evenly distributed to leptons and baryons due to the sphaleron process. Because $\phi$ carries $B-L$ charge $-2$, the final baryon asymmetry will have the same magnitude and opposite sign as the initial $\phi$ asymmetry Eq.~(\ref{eq:Yphi}), i.e. $Y_B=-Y_\phi$.
\\
\section{Majoron phenomenology}

\begin{figure*}
\centering
\includegraphics[width=0.7\textwidth]{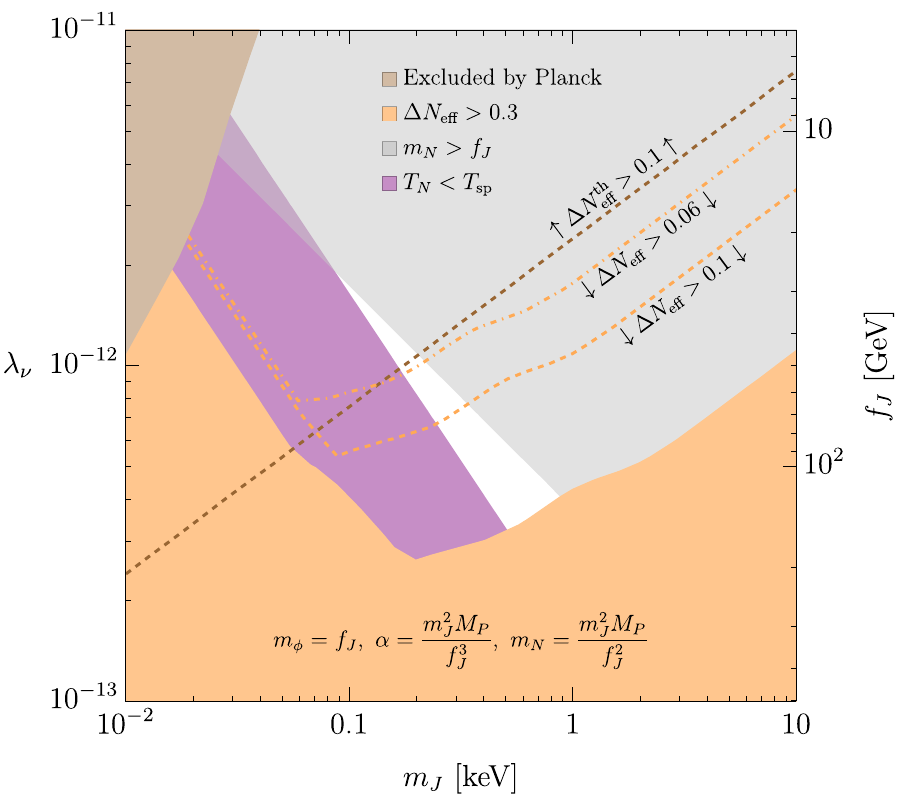}
\caption{Constraints (shaded region) and future sensitivities (dashed and dot-dashed lines) in the $m_J-\lambda_\nu (f_J)$ plane. The brown region is excluded by Planck \cite{Sandner:2023ptm}, and the orange region is excluded from $\Delta N_\textrm{eff}>0.3$ \cite{Planck:2018vyg}.
The constraint $T_N > T_{\rm sp}$ (purple region) gives the upper bound on $f_J$, while the perturbativity condition for $\lambda_N$ (gray region) gives the lower bound.
The allowed parameter space (white region) is near $f_J = \mathcal{O}(100 \GeV)$ and $m_J = \mathcal{O}(0.1-1 \keV)$, which are predicted values by the model.
The dashed and dot-dashed lines are future sensitivities for $\Delta N_\textrm{eff} >0.1$ and $\Delta N_\textrm{eff} >0.06$, respectively.
$\Delta N_\textrm{eff}^\textrm{th}$ indicates the contribution from the thermal production, while $\Delta N_\textrm{eff}$ denotes the contribution from the relic Majoron.
Note all the allowed parameter space can be probed with the sensitivity $\Delta N_\textrm{eff}<0.06$.}
\label{fig:mJlambdaplot}
\end{figure*}

The model has two SM singlet scalar fields: $\phi$ and $\tilde N$, where $\tilde N$ is the superpartner of $N$.
The scalar dynamics at high temperatures is mostly dominated by $\phi$, while $\tilde N$ plays an important role at low temperatures. 
%
From \Eq{eq:model}, one finds the potential for 
$\tilde N$ at low energy scales to be 
\begin{align} \label{eq:RHNpotential}
\Delta V  =&\, (\lambda_N^2|\phi|^2 
\hskip -0.05cm 
+
\hskip -0.05cm  m_{\tilde N}^2
)|\tilde N|^2 + \left(\frac{\alpha \lambda_N m_{\phi}}{2}  \phi \tilde N^2 + h.c. \right)  \nonumber\\ 
& + \frac{\lambda_N^2}{4} |\tilde N|^4 +  {\cal O} \left(\frac{\alpha m_\phi}{M_P} \tilde N^4 \right), 
\end{align}
with $m^2_{\tilde N} ={\cal O}(m^2_\phi)<0$.\footnote{
We assume that the SM singlet scalars
acquire soft SUSY breaking terms
of comparable magnitudes, approximately at the electroweak scale,
i.e.~$m_\phi \sim m_{\tilde{N}}\sim v$. One may consider gauge mediation \cite{Dine:1994vc,Dine:1995ag,Giudice:1998bp}
 or mirage mediation \cite{Choi:2005hd,Choi:2006xb}
to make colored sparticles  
much heavier than $v$  
as required by experimental constraints.
This also allows a wide range of the gravitino mass.} 
Here, the last term is interactions suppressed by $M_P$ and is not relevant for scalar dynamics.
At high temperatures $T > T_{\rm AD}$, $\tilde N $ is trapped at the origin because of a large positive mass contribution from $\lambda_N^2 |\phi|^2 \sim H M_P$.   
After the $Y_\phi$ generation, $\phi$ rolls to the origin with the behavior of $|\phi|^2\propto a^{-3}$. Because of the negative mass squared of $\tilde N$, 
$|\tilde N|$ gradually develops a nonzero expectation value as $\lambda_N^2|\phi|^2$ becomes smaller than $m_{\tilde N}^2$. Thereafter, the asymmetry of  $\phi$ smoothly transfers to that of $\tilde N= \frac{1}{\sqrt{2}}r_N e^{i\theta_N}$. The dynamics of $\tilde N$ conserves the $B-L$ number, and the net density of the AD sector
\beq
\bar n_{B-L}|_{\rm AD} = -2 \bar{n}_\phi + \bar{n}_N = - 2 r_\phi^2 \dot\theta_\phi +  r_N^2  \dot\theta_N 
\equiv f_J^2 \dot\theta_J
\eeq  
is decreasing as $a^{-3}$ until the explicit breaking effect becomes important. The $U(1)_{B-L}$ spontaneous breaking scale $f_J =\sqrt{4 r_\phi^2 +   r_N^2 }$ is gradually dominated by $r_N$. The associated Nambu-Goldstone boson, Majoron, corresponds to $J(x) =   f_J \theta_J(x)$, and the $B-L$ asymmetry is carried by the kinetic energy of $J$ \cite{Chikashige:1980ui,Gelmini:1980re}.

The explicit breaking term of the scalar potential provides a damping effect. 
First of all, it gives a scalar potential for the Majoron as $V_J\sim   m_J^2 f_J^2 \cos(J/f_J)$, 
where 
\beq
m_{J}\sim f_J \sqrt{\frac{\alpha m_\phi}{ \mpl}} = {\cal O}(0.1-1 \keV) \, . 
\eeq 
We note that $\dot\theta_J$ remains much larger than $m_J$ when the AD sector is thermalized with the SM sector, which makes the damping effect negligible so the total $B-L$ number is nearly conserved.
For $T_{\rm sp} \lesssim T\lesssim T_N$, the thermal potential for $r_\phi$ and $r_{N}$ could lead to the trap of the scalar fields at the origin $\langle f_J\rangle =0$ depending on the size of their thermal mass of ${\cal O}(\lambda_N T)$. After the sphaleoron process freezes out,
$Y_B$ is entirely frozen. 

Eventually, $U(1)_{B-L}$ is spontaneously broken with the vacuum expectation values
\begin{align}
&\langle |\phi|\rangle \sim \frac{\alpha m_\phi}{\lambda_N}~~\textrm{and}\\ &\langle  |\tilde N|\rangle \sim f_J \sim \frac{m_\phi}{\lambda_N} \, ,
\end{align}
assuming $|m_{\tilde N}|\sim m_\phi$ and $\alpha \lesssim 1$.

The Majoron emerges after the $B-L$ symmetry is spontaneously broken.
Its mass in our scenario is ${\cal O}(0.1-1)$ keV, which is much smaller than all $B-L$ charged particles except the active neutrinos.
The efficient way to write the relevant Lagrangian for Majoron cosmology is to take a field basis of heavy particles $\Phi_i$ (except the active neutrinos $\nu$) as 
\beq
\Phi_i \to \Phi_i \exp( i q_{\Phi_i} \theta_J)\, ,
\eeq
where $q_{\Phi_i}$ is the $B-L$ charge of $\Phi_i$. Then, only the Majoron and active neutrinos are transformed as 
\beq
U(1)_{B-L} :\quad \theta_J \to \theta_J + c, \quad \nu_i \to e^{-ic}\nu_i.
\eeq 
After integrating out these massive fields, the relevant effective Lagrangian for the Majoron becomes 
\begin{widetext}    
\begin{align}
{\cal L}_{\rm eff} &=\frac{1}{2} f_J^2 (\partial_\mu \theta_J)^2 
- \left(\frac{1}{2} (m_\nu)_{ij} e^{2i\theta_J} \nu_i \nu_j + h.c. \right) \nonumber \\
&\quad 
-\frac{\alpha m}{16M_P}\Big( \kappa\langle r_\phi^4\rangle e^{-8i\theta_J} 
+ \kappa_2\langle r_\phi^3 r_N\rangle e^{-5i\theta_J} 
+ \kappa_3\langle r_\phi^2 r_N^2 \rangle e^{-2i\theta_J} 
+\kappa_4\langle r_\phi r_N^3\rangle e^{i\theta_J}
+\kappa_5\langle r_N^4\rangle e^{4i\theta_J} + h.c.\Big)\\
&=  \frac{1}{2} (\partial_\mu J)^2  - \frac{1}{2} m_J^2 J^2  -  J 
\left(   \frac{i(m_\nu)_{ij}}{f_J} \nu_i\nu_j + h.c.\right) +\cdots ,
\end{align}
\end{widetext}
where $r_\phi =\sqrt{2} |\phi|$, $r_N=\sqrt{2}|\tilde N|$, the Majoron decay constant $f_J = \sqrt{\sum_i 2q_{\Phi_i}^2 \langle |\Phi_i|^2\rangle } = \sqrt{ 4\langle r_\phi^2 \rangle + \langle r_N^2\rangle}$, and the canonically normalized Majoron  $J(x)=f_J \theta_J(x)$. There are several contributions to the Majoron mass from the explicit $B-L$ breaking terms in the superpotential suppressed by $1/M_P$. Although the correct form of the Majoron mass is quite complicated and includes
many parameters, we can easily identify its order of magnitude as $m_J^2 \sim (\alpha m/M_P) f_J^2$.

The Majorons are copiously produced around the $B-L$ phase transition and their density easily becomes an equilibrium value \cite{Li:2023kuz}.
Its abundance is frozen at $T_d\sim 0.1~m_N$, where $m_N$ is the right-handed neutrino mass and $T_d$ is the freeze-out temperature of the Majoron \cite{Escudero:2021rfi}.
Then relativistic Majoron contribution to $\Delta N_\textrm{eff}$ today is given by
\begin{eqnarray}
    \Delta N_\textrm{eff} &=&  \frac{4}{7} \left( \frac{11}{4} 
 \frac{g_{*S}(T_0)}{g_{*S}(T_d)} \right)^{4/3} \, .
\end{eqnarray}
Since the Majoron mass is much smaller than $m_N$, the initial Majorons are relativistic. If the Majoron becomes non-relativistic before it decays away, an additional factor $F_{\rm NR}$ should be included as matter redshifts slower than radiation,
\begin{widetext}
\begin{eqnarray}
    F_{\rm NR} &=& \int \frac{d^3p}{(2\pi)^3} \frac{m_J}{\exp{(p/T_{J,\textrm{decay}})}-1} \bigg / \int \frac{d^3p}{(2\pi)^3} \frac{p}{\exp{(p/T_{J,\textrm{decay}})-1}} \\
    &=& \frac{30\zeta(3)}{\pi^4} \left(
 \frac{g_{*S}(T_0)}{g_{*S}(T_d)} \right)^{-1/3} \frac{m_J}{T_\textrm{decay}} \, ,
\end{eqnarray}
\end{widetext}
where $T_{J,\textrm{decay}}$ and $T_\textrm{decay}$ are the temperatures of the Majoron and the photon, respectively, at which the Majoron decays.
The rate of Majoron decay to neutrinos $\Gamma_J$ is \cite{Akita:2023qiz}
\begin{eqnarray}
    \Gamma_J &=& \frac{m_J}{16 \pi f_J^2} \sum m_\nu^2 \, . 
\end{eqnarray}
By using $\Gamma_J = H$, we get
\begin{eqnarray}
T_\textrm{decay} &\approx& 
     141 \eV \times\\ && \left (\frac{m_J}{1\keV} \right)^{1/2} \left( \frac{100 \GeV}{f_J}\right) \left(\frac{\sum m_\nu^2}{(0.05 \eV)^2}\right)^{1/2} \, , \nonumber 
\end{eqnarray}
assuming the decay happens in the radiation-dominated era. 
For more precise calculations, we need to include the time dilation effect for the lifetime of the Majoron as $\Gamma_J m_J/E_J= H$. However this effect on $\Delta N_{\rm eff}$ is negligible because $\Delta N_\textrm{eff}$ does not depend on $T_\textrm{decay}$ if Majorons decay while relativistic.
The final contribution to $\Delta N_\textrm{eff}$ from the Majoron decays after BBN is given by
\begin{widetext}
\begin{eqnarray}\label{eq:Neff}
    \Delta N_\textrm{eff} &=& 
\frac{4}{7} \left( \frac{11}{4} 
 \frac{g_{*S}(T_0)}{g_{*S}(T_d)} \right)^{4/3} \textrm{Max}[1,F_{\rm NR}]\\
 &\approx& \begin{cases}
 0.029 ~ \left (\frac{100}{g_{*S}(T_d)} \right)^{4/3} &,~ F_{\rm NR}\leq 1\\
 0.23 ~ \left(\frac{m_J}{1 \keV}\right)^{1/2} \left( \frac{f_J}{100 \GeV}\right) \left(\frac{(0.05 \eV)^2}{\sum m_i^2}\right)^{1/2} \left (\frac{100}{g_{*S}(T_d)} \right) &, ~ F_{\rm NR}>1
 \end{cases} \, \, .
 \end{eqnarray}
\end{widetext}
If $f_J$ is excessively large, the Majoron decays much later than when it becomes non-relativistic. Since the energy density of non-relativistic particles redshifts slower than radiation, it contributes more to $N_{\rm eff}$ after decay, as shown in \Eq{eq:Neff}. The $N_{\rm eff}$ bound, $\Delta N_\textrm{eff}<0.3$ ($2\sigma$) from \cite{Planck:2018vyg}, constrains $f_J$ not to significantly exceed the electroweak scale.

In Fig.~\ref{fig:mJlambdaplot}, we show the constrained parameter space along with future $\Delta N_{\rm eff}$ sensitivities in the plane of $m_J$ and $\lambda_\nu (f_J)$, where $\lambda_\nu \approx 0.05 \eV/f_J$ is the Majoron-neutrino coupling. We have used $\mathcal{O}(1)$ model parameters to get $m_J$ and $f_J$ from the model parameters as shown in the plot, so note it can be modified by $\mathcal{O}(1)$ for different parameter choices. A more complete study of the full set of model parameters will be performed in future work.
The orange region and lines come from the $\Delta N_\textrm{eff}$ contribution from the relic Majoron, while the brown region and line are from the late-time production from $J \leftrightarrow \nu \nu$. 
We require $T_N>T_{\rm sp}$, otherwise no baryon asymmetry is generated since thermalization of $N$ happens after the weak sphaleron process ceases.
$m_N<f_J$ is imposed because the RH neutrino mass $m_N = \lambda_N \langle |\phi| \rangle$ needs to be smaller than the Majoron decay constant $f_J>\langle |\phi| \rangle$ with $\lambda_N<{\cal O}(1)$.
The allowed parameter space nearly points $f_J ={\cal O}(100 \GeV)$ and $m_J={\cal O}(0.1 - 1 \keV)$, which can be well predicted by our scenario. The future cosmic microwave background (CMB) observations for $\Delta N_{\rm eff}$ from Simons Observatory \cite{SimonsObservatory:2018koc} or CMB-S4 \cite{CMB-S4:2016ple} can probe all the allowed parameter space. 
\\

\section{Discussion}
The neutrino-portal Affleck-Dine mechanism yields the observed small baryon asymmetry in the Universe as a direct consequence of a hierarchy between two mass parameters inherent in the scenario, $m_\phi$ and $M_P$. Both parameters are linked to the symmetry-breaking scales and provide an organizing principle of the scalar potential.
$M_P$ emerges as a cutoff scale in higher-dimensional operators that break global symmetries due to the quantum gravity effects.
On the other hand, $m_\phi$ represents the soft SUSY breaking scale, which is the energy scale beyond which the potential of a scalar field remains shielded against UV-sensitive quantum corrections.
This naturally prompts the exploration of the correlation between $m_\phi$ and the electroweak scale $v$.

An interesting relation between $m_\phi$ and $v$ arises in phenomenological observables as well. As shown in \Fig{fig:mJlambdaplot}, the phenomenologically allowed values for the Majoron decay constant $f_J$, hence also for $m_\phi$, are close to the electroweak scale. This result is independent of the theoretical motivation for linking $m_\phi$ to $v$. 
Our scenario establishes a fundamental correlation between the electroweak scale and the origin of asymmetry in the Universe. 

In contrast to conventional high-scale baryogenesis models, which typically do not predict any observable for new physics, our model has distinct low-energy observable implications in cosmology, yet asymmetry is still generated at a high scale, $T_{\rm AD}\sim 10^{10}\GeV$. 
The presence of a light Majoron with a keV mass is well predicted in the model, and the consequential effects on $N_{\rm eff}$ can be measured in the near future.

The reheating temperature needs to be higher than $T_{\rm AD}$ for the AD mechanism to work, but it cannot be significantly higher for two reasons. 
Firstly, a higher reheating temperature leads to a larger baryon isocurvature perturbation induced by a light field. 
During inflation with the Hubble rate $H_I$, the angular field has a perturbation as $\delta\phi \sim H_I/2\pi$, 
leading to
$\delta\phi/\langle\phi\rangle \sim 0.1\sqrt{H_I/M_P}$. 
From the bound on isocurvature perturbation \cite{Planck:2018jri}, $\delta\phi/\langle\phi\rangle \lesssim 10^{-6}$.
This gives an upper bound on the reheating temperature $T_R$ as $T_R\lesssim 10^{-5}M_P \sim 10^{13}\,{\rm GeV}$.
Secondly, the scalar field $\phi$ undergoes negative damping at temperatures above $T_{\rm AD}$ if it is displaced from the fixed point \cite{Dine:1995kz,Harigaya:2015hha}. 
The AD mechanism occurring shortly after reheating successfully exhibits all the previously discussed properties without these issues.

Our minimal scenario with SUSY has another phenomenological observable, which is that the lightest neutrino is almost massless, $m_{\rm light}\sim (m_N/M_P) \sum m_\nu$. 
This can explain the smallness of the neutrino mass sum derived from the recent DESI data analysis \cite{DESI:2024mwx}.
The more detailed connection to the spectrum of superpartners of the SM particles is also an important question, and it requires a more dedicated study of the SUSY-breaking mediation mechanism that will be revealed from the future observation of SUSY particles \cite{ParticleDataGroup:2022pth,Allanach:2024suz}.
We leave this aspect to future works because all other aspects presented in this work yield consistent results as long as the same scalar potentials are employed. Here, SUSY only serves as a tool for organizing the scalar potentials.
\\
\acknowledgments{
We are grateful to Arushi Bodas and Keisuke Harigaya for helpful comments. JHC is supported by Fermi Research Alliance, LLC under Contract No. DE-AC02-07CH11359 with the U.S. Department of Energy, Office of Science, Office of High Energy Physics. KSJ is supported by the National Research Foundation (NRF) of Korea grants funded by the Korea
government: Grants No.~2021R1A4A5031460 and RS-2023-00249330. CHL and CSS are also supported by the NRF of Korea (NRF-2022R1C1C1011840, NRF-2022R1A4A5030362). This work was performed in part at the workshop ``Dark Matter as a Portal to New Physics" supported by APCTP.}

\bibliography{ref}

\begin{thebibliography}{50}%
\makeatletter
\providecommand \@ifxundefined [1]{%
 \@ifx{#1\undefined}
}%
\providecommand \@ifnum [1]{%
 \ifnum #1\expandafter \@firstoftwo
 \else \expandafter \@secondoftwo
 \fi
}%
\providecommand \@ifx [1]{%
 \ifx #1\expandafter \@firstoftwo
 \else \expandafter \@secondoftwo
 \fi
}%
\providecommand \natexlab [1]{#1}%
\providecommand \enquote  [1]{``#1''}%
\providecommand \bibnamefont  [1]{#1}%
\providecommand \bibfnamefont [1]{#1}%
\providecommand \citenamefont [1]{#1}%
\providecommand \href@noop [0]{\@secondoftwo}%
\providecommand \href [0]{\begingroup \@sanitize@url \@href}%
\providecommand \@href[1]{\@@startlink{#1}\@@href}%
\providecommand \@@href[1]{\endgroup#1\@@endlink}%
\providecommand \@sanitize@url [0]{\catcode `\\12\catcode `\$12\catcode
  `\&12\catcode `\#12\catcode `\^12\catcode `\_12\catcode `\%12\relax}%
\providecommand \@@startlink[1]{}%
\providecommand \@@endlink[0]{}%
\providecommand \url  [0]{\begingroup\@sanitize@url \@url }%
\providecommand \@url [1]{\endgroup\@href {#1}{\urlprefix }}%
\providecommand \urlprefix  [0]{URL }%
\providecommand \Eprint [0]{\href }%
\providecommand \doibase [0]{http://dx.doi.org/}%
\providecommand \selectlanguage [0]{\@gobble}%
\providecommand \bibinfo  [0]{\@secondoftwo}%
\providecommand \bibfield  [0]{\@secondoftwo}%
\providecommand \translation [1]{[#1]}%
\providecommand \BibitemOpen [0]{}%
\providecommand \bibitemStop [0]{}%
\providecommand \bibitemNoStop [0]{.\EOS\space}%
\providecommand \EOS [0]{\spacefactor3000\relax}%
\providecommand \BibitemShut  [1]{\csname bibitem#1\endcsname}%
\let\auto@bib@innerbib\@empty
\bibitem [{\citenamefont {Aghanim}\ \emph {et~al.}(2020)\citenamefont {Aghanim}
  \emph {et~al.}}]{Planck:2018vyg}%
  \BibitemOpen
  \bibfield  {author} {\bibinfo {author} {\bibfnamefont {N.}~\bibnamefont
  {Aghanim}} \emph {et~al.} (\bibinfo {collaboration} {Planck}),\ }\href
  {\doibase 10.1051/0004-6361/201833910} {\bibfield  {journal} {\bibinfo
  {journal} {Astron. Astrophys.}\ }\textbf {\bibinfo {volume} {641}},\ \bibinfo
  {pages} {A6} (\bibinfo {year} {2020})},\ \bibinfo {note} {[Erratum:
  Astron.Astrophys. 652, C4 (2021)]},\ \Eprint
  {http://arxiv.org/abs/1807.06209} {arXiv:1807.06209 [astro-ph.CO]}
  \BibitemShut {NoStop}%
\bibitem [{\citenamefont {Bodeker}\ and\ \citenamefont
  {Buchmuller}(2021)}]{Bodeker:2020ghk}%
  \BibitemOpen
  \bibfield  {author} {\bibinfo {author} {\bibfnamefont {D.}~\bibnamefont
  {Bodeker}}\ and\ \bibinfo {author} {\bibfnamefont {W.}~\bibnamefont
  {Buchmuller}},\ }\href {\doibase 10.1103/RevModPhys.93.035004} {\bibfield
  {journal} {\bibinfo  {journal} {Rev. Mod. Phys.}\ }\textbf {\bibinfo {volume}
  {93}},\ \bibinfo {pages} {035004} (\bibinfo {year} {2021})},\ \Eprint
  {http://arxiv.org/abs/2009.07294} {arXiv:2009.07294 [hep-ph]} \BibitemShut
  {NoStop}%
\bibitem [{\citenamefont {Elor}\ \emph {et~al.}(2022)\citenamefont {Elor} \emph
  {et~al.}}]{Elor:2022hpa}%
  \BibitemOpen
  \bibfield  {author} {\bibinfo {author} {\bibfnamefont {G.}~\bibnamefont
  {Elor}} \emph {et~al.},\ }in\ \href@noop {} {\emph {\bibinfo {booktitle}
  {{Snowmass 2021}}}}\ (\bibinfo {year} {2022})\ \Eprint
  {http://arxiv.org/abs/2203.05010} {arXiv:2203.05010 [hep-ph]} \BibitemShut
  {NoStop}%
\bibitem [{\citenamefont {Barrow}\ \emph {et~al.}(2022)\citenamefont {Barrow}
  \emph {et~al.}}]{Barrow:2022gsu}%
  \BibitemOpen
  \bibfield  {author} {\bibinfo {author} {\bibfnamefont {J.~L.}\ \bibnamefont
  {Barrow}} \emph {et~al.},\ }\href@noop {} {\  (\bibinfo {year} {2022})},\
  \Eprint {http://arxiv.org/abs/2203.07059} {arXiv:2203.07059 [hep-ph]}
  \BibitemShut {NoStop}%
\bibitem [{\citenamefont {Affleck}\ and\ \citenamefont
  {Dine}(1985)}]{Affleck:1984fy}%
  \BibitemOpen
  \bibfield  {author} {\bibinfo {author} {\bibfnamefont {I.}~\bibnamefont
  {Affleck}}\ and\ \bibinfo {author} {\bibfnamefont {M.}~\bibnamefont {Dine}},\
  }\href {\doibase 10.1016/0550-3213(85)90021-5} {\bibfield  {journal}
  {\bibinfo  {journal} {Nucl. Phys. B}\ }\textbf {\bibinfo {volume} {249}},\
  \bibinfo {pages} {361} (\bibinfo {year} {1985})}\BibitemShut {NoStop}%
\bibitem [{\citenamefont {Dine}\ \emph
  {et~al.}(1996{\natexlab{a}})\citenamefont {Dine}, \citenamefont {Randall},\
  and\ \citenamefont {Thomas}}]{Dine:1995kz}%
  \BibitemOpen
  \bibfield  {author} {\bibinfo {author} {\bibfnamefont {M.}~\bibnamefont
  {Dine}}, \bibinfo {author} {\bibfnamefont {L.}~\bibnamefont {Randall}}, \
  and\ \bibinfo {author} {\bibfnamefont {S.~D.}\ \bibnamefont {Thomas}},\
  }\href {\doibase 10.1016/0550-3213(95)00538-2} {\bibfield  {journal}
  {\bibinfo  {journal} {Nucl. Phys. B}\ }\textbf {\bibinfo {volume} {458}},\
  \bibinfo {pages} {291} (\bibinfo {year} {1996}{\natexlab{a}})},\ \Eprint
  {http://arxiv.org/abs/hep-ph/9507453} {arXiv:hep-ph/9507453} \BibitemShut
  {NoStop}%
\bibitem [{\citenamefont {Allahverdi}\ and\ \citenamefont
  {Mazumdar}(2012)}]{Allahverdi:2012ju}%
  \BibitemOpen
  \bibfield  {author} {\bibinfo {author} {\bibfnamefont {R.}~\bibnamefont
  {Allahverdi}}\ and\ \bibinfo {author} {\bibfnamefont {A.}~\bibnamefont
  {Mazumdar}},\ }\href {\doibase 10.1088/1367-2630/14/12/125013} {\bibfield
  {journal} {\bibinfo  {journal} {New J. Phys.}\ }\textbf {\bibinfo {volume}
  {14}},\ \bibinfo {pages} {125013} (\bibinfo {year} {2012})}\BibitemShut
  {NoStop}%
\bibitem [{sus(2000)}]{susy1:KaneShifman}%
  \BibitemOpen
  \href@noop {} {\emph {\bibinfo {title} {The Supersymmetric World---The
  Beginnings of the Theory}}}\ (\bibinfo  {publisher} {World Scientific},\
  \bibinfo {address} {Singapore},\ \bibinfo {year} {2000})\ \bibinfo {note}
  {edited by G.~Kane and M.~Shifman, contains an early history of supersymmetry
  and a guide to the original literature}\BibitemShut {NoStop}%
\bibitem [{\citenamefont {Ellis}\ \emph {et~al.}(1984)\citenamefont {Ellis},
  \citenamefont {Hagelin}, \citenamefont {Nanopoulos}, \citenamefont {Olive},\
  and\ \citenamefont {Srednicki}}]{Ellis:1983ew}%
  \BibitemOpen
  \bibfield  {author} {\bibinfo {author} {\bibfnamefont {J.~R.}\ \bibnamefont
  {Ellis}}, \bibinfo {author} {\bibfnamefont {J.~S.}\ \bibnamefont {Hagelin}},
  \bibinfo {author} {\bibfnamefont {D.~V.}\ \bibnamefont {Nanopoulos}},
  \bibinfo {author} {\bibfnamefont {K.~A.}\ \bibnamefont {Olive}}, \ and\
  \bibinfo {author} {\bibfnamefont {M.}~\bibnamefont {Srednicki}},\ }\bibfield
  {booktitle} {\emph {\bibinfo {booktitle} {{Particle physics and cosmology:
  Dark matter}}},\ }\href {\doibase 10.1016/0550-3213(84)90461-9} {\bibfield
  {journal} {\bibinfo  {journal} {Nucl. Phys.}\ }\textbf {\bibinfo {volume}
  {B238}},\ \bibinfo {pages} {453} (\bibinfo {year} {1984})}\BibitemShut
  {NoStop}%
\bibitem [{\citenamefont {Nilles}\ \emph {et~al.}(1983)\citenamefont {Nilles},
  \citenamefont {Srednicki},\ and\ \citenamefont {Wyler}}]{Nilles:1982dy}%
  \BibitemOpen
  \bibfield  {author} {\bibinfo {author} {\bibfnamefont {H.~P.}\ \bibnamefont
  {Nilles}}, \bibinfo {author} {\bibfnamefont {M.}~\bibnamefont {Srednicki}}, \
  and\ \bibinfo {author} {\bibfnamefont {D.}~\bibnamefont {Wyler}},\ }\href
  {\doibase 10.1016/0370-2693(83)90460-4} {\bibfield  {journal} {\bibinfo
  {journal} {Phys. Lett.}\ }\textbf {\bibinfo {volume} {120B}},\ \bibinfo
  {pages} {346} (\bibinfo {year} {1983})}\BibitemShut {NoStop}%
\bibitem [{\citenamefont {Salam}\ and\ \citenamefont
  {Strathdee}(1974)}]{Salam:1974yz}%
  \BibitemOpen
  \bibfield  {author} {\bibinfo {author} {\bibfnamefont {A.}~\bibnamefont
  {Salam}}\ and\ \bibinfo {author} {\bibfnamefont {J.~A.}\ \bibnamefont
  {Strathdee}},\ }\href {\doibase 10.1016/0550-3213(74)90537-9} {\bibfield
  {journal} {\bibinfo  {journal} {Nucl. Phys.}\ }\textbf {\bibinfo {volume}
  {B76}},\ \bibinfo {pages} {477} (\bibinfo {year} {1974})}\BibitemShut
  {NoStop}%
\bibitem [{\citenamefont {Minkowski}(1977)}]{Minkowski:1977sc}%
  \BibitemOpen
  \bibfield  {author} {\bibinfo {author} {\bibfnamefont {P.}~\bibnamefont
  {Minkowski}},\ }\href {\doibase 10.1016/0370-2693(77)90435-X} {\bibfield
  {journal} {\bibinfo  {journal} {Phys. Lett. B}\ }\textbf {\bibinfo {volume}
  {67}},\ \bibinfo {pages} {421} (\bibinfo {year} {1977})}\BibitemShut
  {NoStop}%
\bibitem [{\citenamefont {Ramond}(1979)}]{Ramond:1979py}%
  \BibitemOpen
  \bibfield  {author} {\bibinfo {author} {\bibfnamefont {P.}~\bibnamefont
  {Ramond}},\ }in\ \href@noop {} {\emph {\bibinfo {booktitle} {{International
  Symposium on Fundamentals of Quantum Theory and Quantum Field Theory}}}}\
  (\bibinfo {year} {1979})\ \Eprint {http://arxiv.org/abs/hep-ph/9809459}
  {arXiv:hep-ph/9809459} \BibitemShut {NoStop}%
\bibitem [{\citenamefont {Gell-Mann}\ \emph {et~al.}(1979)\citenamefont
  {Gell-Mann}, \citenamefont {Ramond},\ and\ \citenamefont
  {Slansky}}]{Gell-Mann:1979vob}%
  \BibitemOpen
  \bibfield  {author} {\bibinfo {author} {\bibfnamefont {M.}~\bibnamefont
  {Gell-Mann}}, \bibinfo {author} {\bibfnamefont {P.}~\bibnamefont {Ramond}}, \
  and\ \bibinfo {author} {\bibfnamefont {R.}~\bibnamefont {Slansky}},\
  }\href@noop {} {\bibfield  {journal} {\bibinfo  {journal} {Conf. Proc. C}\
  }\textbf {\bibinfo {volume} {790927}},\ \bibinfo {pages} {315} (\bibinfo
  {year} {1979})},\ \Eprint {http://arxiv.org/abs/1306.4669} {arXiv:1306.4669
  [hep-th]} \BibitemShut {NoStop}%
\bibitem [{\citenamefont {Yanagida}(1979)}]{Yanagida:1979as}%
  \BibitemOpen
  \bibfield  {author} {\bibinfo {author} {\bibfnamefont {T.}~\bibnamefont
  {Yanagida}},\ }\href@noop {} {\bibfield  {journal} {\bibinfo  {journal}
  {Conf. Proc. C}\ }\textbf {\bibinfo {volume} {7902131}},\ \bibinfo {pages}
  {95} (\bibinfo {year} {1979})}\BibitemShut {NoStop}%
\bibitem [{\citenamefont {Mohapatra}\ and\ \citenamefont
  {Senjanovic}(1981)}]{Mohapatra:1980yp}%
  \BibitemOpen
  \bibfield  {author} {\bibinfo {author} {\bibfnamefont {R.~N.}\ \bibnamefont
  {Mohapatra}}\ and\ \bibinfo {author} {\bibfnamefont {G.}~\bibnamefont
  {Senjanovic}},\ }\href {\doibase 10.1103/PhysRevD.23.165} {\bibfield
  {journal} {\bibinfo  {journal} {Phys. Rev. D}\ }\textbf {\bibinfo {volume}
  {23}},\ \bibinfo {pages} {165} (\bibinfo {year} {1981})}\BibitemShut
  {NoStop}%
\bibitem [{\citenamefont {Holman}\ \emph {et~al.}(1992)\citenamefont {Holman},
  \citenamefont {Hsu}, \citenamefont {Kephart}, \citenamefont {Kolb},
  \citenamefont {Watkins},\ and\ \citenamefont {Widrow}}]{Holman:1992us}%
  \BibitemOpen
  \bibfield  {author} {\bibinfo {author} {\bibfnamefont {R.}~\bibnamefont
  {Holman}}, \bibinfo {author} {\bibfnamefont {S.~D.~H.}\ \bibnamefont {Hsu}},
  \bibinfo {author} {\bibfnamefont {T.~W.}\ \bibnamefont {Kephart}}, \bibinfo
  {author} {\bibfnamefont {E.~W.}\ \bibnamefont {Kolb}}, \bibinfo {author}
  {\bibfnamefont {R.}~\bibnamefont {Watkins}}, \ and\ \bibinfo {author}
  {\bibfnamefont {L.~M.}\ \bibnamefont {Widrow}},\ }\href {\doibase
  10.1016/0370-2693(92)90491-L} {\bibfield  {journal} {\bibinfo  {journal}
  {Phys. Lett. B}\ }\textbf {\bibinfo {volume} {282}},\ \bibinfo {pages} {132}
  (\bibinfo {year} {1992})},\ \Eprint {http://arxiv.org/abs/hep-ph/9203206}
  {arXiv:hep-ph/9203206} \BibitemShut {NoStop}%
\bibitem [{\citenamefont {Kamionkowski}\ and\ \citenamefont
  {March-Russell}(1992)}]{Kamionkowski:1992mf}%
  \BibitemOpen
  \bibfield  {author} {\bibinfo {author} {\bibfnamefont {M.}~\bibnamefont
  {Kamionkowski}}\ and\ \bibinfo {author} {\bibfnamefont {J.}~\bibnamefont
  {March-Russell}},\ }\href {\doibase 10.1016/0370-2693(92)90492-M} {\bibfield
  {journal} {\bibinfo  {journal} {Phys. Lett. B}\ }\textbf {\bibinfo {volume}
  {282}},\ \bibinfo {pages} {137} (\bibinfo {year} {1992})},\ \Eprint
  {http://arxiv.org/abs/hep-th/9202003} {arXiv:hep-th/9202003} \BibitemShut
  {NoStop}%
\bibitem [{\citenamefont {Barr}\ and\ \citenamefont
  {Seckel}(1992)}]{Barr:1992qq}%
  \BibitemOpen
  \bibfield  {author} {\bibinfo {author} {\bibfnamefont {S.~M.}\ \bibnamefont
  {Barr}}\ and\ \bibinfo {author} {\bibfnamefont {D.}~\bibnamefont {Seckel}},\
  }\href {\doibase 10.1103/PhysRevD.46.539} {\bibfield  {journal} {\bibinfo
  {journal} {Phys. Rev. D}\ }\textbf {\bibinfo {volume} {46}},\ \bibinfo
  {pages} {539} (\bibinfo {year} {1992})}\BibitemShut {NoStop}%
\bibitem [{\citenamefont {Witten}(1981)}]{Witten:1981nf}%
  \BibitemOpen
  \bibfield  {author} {\bibinfo {author} {\bibfnamefont {E.}~\bibnamefont
  {Witten}},\ }\href {\doibase 10.1016/0550-3213(81)90006-7} {\bibfield
  {journal} {\bibinfo  {journal} {Nucl. Phys.}\ }\textbf {\bibinfo {volume}
  {B188}},\ \bibinfo {pages} {513} (\bibinfo {year} {1981})}\BibitemShut
  {NoStop}%
\bibitem [{\citenamefont {Dimopoulos}\ and\ \citenamefont
  {Georgi}(1981)}]{Dimopoulos:1981zb}%
  \BibitemOpen
  \bibfield  {author} {\bibinfo {author} {\bibfnamefont {S.}~\bibnamefont
  {Dimopoulos}}\ and\ \bibinfo {author} {\bibfnamefont {H.}~\bibnamefont
  {Georgi}},\ }\href {\doibase 10.1016/0550-3213(81)90522-8} {\bibfield
  {journal} {\bibinfo  {journal} {Nucl. Phys.}\ }\textbf {\bibinfo {volume}
  {B193}},\ \bibinfo {pages} {150} (\bibinfo {year} {1981})}\BibitemShut
  {NoStop}%
\bibitem [{\citenamefont {Susskind}(1984)}]{Susskind:1982mw}%
  \BibitemOpen
  \bibfield  {author} {\bibinfo {author} {\bibfnamefont {L.}~\bibnamefont
  {Susskind}},\ }\href {\doibase 10.1016/0370-1573(84)90208-4} {\bibfield
  {journal} {\bibinfo  {journal} {Phys. Rept.}\ }\textbf {\bibinfo {volume}
  {104}},\ \bibinfo {pages} {181} (\bibinfo {year} {1984})}\BibitemShut
  {NoStop}%
\bibitem [{\citenamefont {Girardello}\ and\ \citenamefont
  {Grisaru}(1982)}]{Girardello:1981wz}%
  \BibitemOpen
  \bibfield  {author} {\bibinfo {author} {\bibfnamefont {L.}~\bibnamefont
  {Girardello}}\ and\ \bibinfo {author} {\bibfnamefont {M.~T.}\ \bibnamefont
  {Grisaru}},\ }\href {\doibase 10.1016/0550-3213(82)90512-0} {\bibfield
  {journal} {\bibinfo  {journal} {Nucl. Phys.}\ }\textbf {\bibinfo {volume}
  {B194}},\ \bibinfo {pages} {65} (\bibinfo {year} {1982})}\BibitemShut
  {NoStop}%
\bibitem [{\citenamefont {Hall}\ and\ \citenamefont
  {Randall}(1990)}]{Hall:1990ac}%
  \BibitemOpen
  \bibfield  {author} {\bibinfo {author} {\bibfnamefont {L.~J.}\ \bibnamefont
  {Hall}}\ and\ \bibinfo {author} {\bibfnamefont {L.}~\bibnamefont {Randall}},\
  }\href {\doibase 10.1103/PhysRevLett.65.2939} {\bibfield  {journal} {\bibinfo
   {journal} {Phys. Rev. Lett.}\ }\textbf {\bibinfo {volume} {65}},\ \bibinfo
  {pages} {2939} (\bibinfo {year} {1990})}\BibitemShut {NoStop}%
\bibitem [{\citenamefont {Kawasaki}\ and\ \citenamefont
  {Takesako}(2012)}]{Kawasaki:2011zi}%
  \BibitemOpen
  \bibfield  {author} {\bibinfo {author} {\bibfnamefont {M.}~\bibnamefont
  {Kawasaki}}\ and\ \bibinfo {author} {\bibfnamefont {T.}~\bibnamefont
  {Takesako}},\ }\href {\doibase 10.1016/j.physletb.2012.03.069} {\bibfield
  {journal} {\bibinfo  {journal} {Phys. Lett. B}\ }\textbf {\bibinfo {volume}
  {711}},\ \bibinfo {pages} {173} (\bibinfo {year} {2012})},\ \Eprint
  {http://arxiv.org/abs/1112.5823} {arXiv:1112.5823 [hep-ph]} \BibitemShut
  {NoStop}%
\bibitem [{\citenamefont {Sakharov}(1967)}]{Sakharov:1967dj}%
  \BibitemOpen
  \bibfield  {author} {\bibinfo {author} {\bibfnamefont {A.~D.}\ \bibnamefont
  {Sakharov}},\ }\href {\doibase 10.1070/PU1991v034n05ABEH002497} {\bibfield
  {journal} {\bibinfo  {journal} {Pisma Zh. Eksp. Teor. Fiz.}\ }\textbf
  {\bibinfo {volume} {5}},\ \bibinfo {pages} {32} (\bibinfo {year}
  {1967})}\BibitemShut {NoStop}%
\bibitem [{\citenamefont {Besak}\ and\ \citenamefont
  {Bodeker}(2012)}]{Besak:2012qm}%
  \BibitemOpen
  \bibfield  {author} {\bibinfo {author} {\bibfnamefont {D.}~\bibnamefont
  {Besak}}\ and\ \bibinfo {author} {\bibfnamefont {D.}~\bibnamefont
  {Bodeker}},\ }\href {\doibase 10.1088/1475-7516/2012/03/029} {\bibfield
  {journal} {\bibinfo  {journal} {JCAP}\ }\textbf {\bibinfo {volume} {03}},\
  \bibinfo {pages} {029} (\bibinfo {year} {2012})},\ \Eprint
  {http://arxiv.org/abs/1202.1288} {arXiv:1202.1288 [hep-ph]} \BibitemShut
  {NoStop}%
\bibitem [{\citenamefont {Garbrecht}\ \emph {et~al.}(2013)\citenamefont
  {Garbrecht}, \citenamefont {Glowna},\ and\ \citenamefont
  {Schwaller}}]{Garbrecht:2013bia}%
  \BibitemOpen
  \bibfield  {author} {\bibinfo {author} {\bibfnamefont {B.}~\bibnamefont
  {Garbrecht}}, \bibinfo {author} {\bibfnamefont {F.}~\bibnamefont {Glowna}}, \
  and\ \bibinfo {author} {\bibfnamefont {P.}~\bibnamefont {Schwaller}},\ }\href
  {\doibase 10.1016/j.nuclphysb.2013.08.020} {\bibfield  {journal} {\bibinfo
  {journal} {Nucl. Phys. B}\ }\textbf {\bibinfo {volume} {877}},\ \bibinfo
  {pages} {1} (\bibinfo {year} {2013})},\ \Eprint
  {http://arxiv.org/abs/1303.5498} {arXiv:1303.5498 [hep-ph]} \BibitemShut
  {NoStop}%
\bibitem [{\citenamefont {Ghisoiu}\ and\ \citenamefont
  {Laine}(2014)}]{Ghisoiu:2014ena}%
  \BibitemOpen
  \bibfield  {author} {\bibinfo {author} {\bibfnamefont {I.}~\bibnamefont
  {Ghisoiu}}\ and\ \bibinfo {author} {\bibfnamefont {M.}~\bibnamefont
  {Laine}},\ }\href {\doibase 10.1088/1475-7516/2014/12/032} {\bibfield
  {journal} {\bibinfo  {journal} {JCAP}\ }\textbf {\bibinfo {volume} {12}},\
  \bibinfo {pages} {032} (\bibinfo {year} {2014})},\ \Eprint
  {http://arxiv.org/abs/1411.1765} {arXiv:1411.1765 [hep-ph]} \BibitemShut
  {NoStop}%
\bibitem [{\citenamefont {Escudero}\ and\ \citenamefont
  {Witte}(2020)}]{Escudero:2019gvw}%
  \BibitemOpen
  \bibfield  {author} {\bibinfo {author} {\bibfnamefont {M.}~\bibnamefont
  {Escudero}}\ and\ \bibinfo {author} {\bibfnamefont {S.~J.}\ \bibnamefont
  {Witte}},\ }\href {\doibase 10.1140/epjc/s10052-020-7854-5} {\bibfield
  {journal} {\bibinfo  {journal} {Eur. Phys. J. C}\ }\textbf {\bibinfo {volume}
  {80}},\ \bibinfo {pages} {294} (\bibinfo {year} {2020})},\ \Eprint
  {http://arxiv.org/abs/1909.04044} {arXiv:1909.04044 [astro-ph.CO]}
  \BibitemShut {NoStop}%
\bibitem [{\citenamefont {Anisimov}\ and\ \citenamefont
  {Dine}(2001)}]{Anisimov:2000wx}%
  \BibitemOpen
  \bibfield  {author} {\bibinfo {author} {\bibfnamefont {A.}~\bibnamefont
  {Anisimov}}\ and\ \bibinfo {author} {\bibfnamefont {M.}~\bibnamefont
  {Dine}},\ }\href {\doibase 10.1016/S0550-3213(01)00550-8} {\bibfield
  {journal} {\bibinfo  {journal} {Nucl. Phys. B}\ }\textbf {\bibinfo {volume}
  {619}},\ \bibinfo {pages} {729} (\bibinfo {year} {2001})},\ \Eprint
  {http://arxiv.org/abs/hep-ph/0008058} {arXiv:hep-ph/0008058} \BibitemShut
  {NoStop}%
\bibitem [{\citenamefont {D'Onofrio}\ \emph {et~al.}(2014)\citenamefont
  {D'Onofrio}, \citenamefont {Rummukainen},\ and\ \citenamefont
  {Tranberg}}]{DOnofrio:2014rug}%
  \BibitemOpen
  \bibfield  {author} {\bibinfo {author} {\bibfnamefont {M.}~\bibnamefont
  {D'Onofrio}}, \bibinfo {author} {\bibfnamefont {K.}~\bibnamefont
  {Rummukainen}}, \ and\ \bibinfo {author} {\bibfnamefont {A.}~\bibnamefont
  {Tranberg}},\ }\href {\doibase 10.1103/PhysRevLett.113.141602} {\bibfield
  {journal} {\bibinfo  {journal} {Phys. Rev. Lett.}\ }\textbf {\bibinfo
  {volume} {113}},\ \bibinfo {pages} {141602} (\bibinfo {year} {2014})},\
  \Eprint {http://arxiv.org/abs/1404.3565} {arXiv:1404.3565 [hep-ph]}
  \BibitemShut {NoStop}%
\bibitem [{\citenamefont {Sandner}\ \emph {et~al.}(2023)\citenamefont
  {Sandner}, \citenamefont {Escudero},\ and\ \citenamefont
  {Witte}}]{Sandner:2023ptm}%
  \BibitemOpen
  \bibfield  {author} {\bibinfo {author} {\bibfnamefont {S.}~\bibnamefont
  {Sandner}}, \bibinfo {author} {\bibfnamefont {M.}~\bibnamefont {Escudero}}, \
  and\ \bibinfo {author} {\bibfnamefont {S.~J.}\ \bibnamefont {Witte}},\ }\href
  {\doibase 10.1140/epjc/s10052-023-11864-6} {\bibfield  {journal} {\bibinfo
  {journal} {Eur. Phys. J. C}\ }\textbf {\bibinfo {volume} {83}},\ \bibinfo
  {pages} {709} (\bibinfo {year} {2023})},\ \Eprint
  {http://arxiv.org/abs/2305.01692} {arXiv:2305.01692 [hep-ph]} \BibitemShut
  {NoStop}%
\bibitem [{\citenamefont {Dine}\ \emph {et~al.}(1995)\citenamefont {Dine},
  \citenamefont {Nelson},\ and\ \citenamefont {Shirman}}]{Dine:1994vc}%
  \BibitemOpen
  \bibfield  {author} {\bibinfo {author} {\bibfnamefont {M.}~\bibnamefont
  {Dine}}, \bibinfo {author} {\bibfnamefont {A.~E.}\ \bibnamefont {Nelson}}, \
  and\ \bibinfo {author} {\bibfnamefont {Y.}~\bibnamefont {Shirman}},\ }\href
  {\doibase 10.1103/PhysRevD.51.1362} {\bibfield  {journal} {\bibinfo
  {journal} {Phys. Rev. D}\ }\textbf {\bibinfo {volume} {51}},\ \bibinfo
  {pages} {1362} (\bibinfo {year} {1995})},\ \Eprint
  {http://arxiv.org/abs/hep-ph/9408384} {arXiv:hep-ph/9408384} \BibitemShut
  {NoStop}%
\bibitem [{\citenamefont {Dine}\ \emph
  {et~al.}(1996{\natexlab{b}})\citenamefont {Dine}, \citenamefont {Nelson},
  \citenamefont {Nir},\ and\ \citenamefont {Shirman}}]{Dine:1995ag}%
  \BibitemOpen
  \bibfield  {author} {\bibinfo {author} {\bibfnamefont {M.}~\bibnamefont
  {Dine}}, \bibinfo {author} {\bibfnamefont {A.~E.}\ \bibnamefont {Nelson}},
  \bibinfo {author} {\bibfnamefont {Y.}~\bibnamefont {Nir}}, \ and\ \bibinfo
  {author} {\bibfnamefont {Y.}~\bibnamefont {Shirman}},\ }\href {\doibase
  10.1103/PhysRevD.53.2658} {\bibfield  {journal} {\bibinfo  {journal} {Phys.
  Rev. D}\ }\textbf {\bibinfo {volume} {53}},\ \bibinfo {pages} {2658}
  (\bibinfo {year} {1996}{\natexlab{b}})},\ \Eprint
  {http://arxiv.org/abs/hep-ph/9507378} {arXiv:hep-ph/9507378} \BibitemShut
  {NoStop}%
\bibitem [{\citenamefont {Giudice}\ and\ \citenamefont
  {Rattazzi}(1999)}]{Giudice:1998bp}%
  \BibitemOpen
  \bibfield  {author} {\bibinfo {author} {\bibfnamefont {G.~F.}\ \bibnamefont
  {Giudice}}\ and\ \bibinfo {author} {\bibfnamefont {R.}~\bibnamefont
  {Rattazzi}},\ }\href {\doibase 10.1016/S0370-1573(99)00042-3} {\bibfield
  {journal} {\bibinfo  {journal} {Phys. Rept.}\ }\textbf {\bibinfo {volume}
  {322}},\ \bibinfo {pages} {419} (\bibinfo {year} {1999})},\ \Eprint
  {http://arxiv.org/abs/hep-ph/9801271} {arXiv:hep-ph/9801271} \BibitemShut
  {NoStop}%
\bibitem [{\citenamefont {Choi}\ \emph {et~al.}(2006)\citenamefont {Choi},
  \citenamefont {Jeong}, \citenamefont {Kobayashi},\ and\ \citenamefont
  {Okumura}}]{Choi:2005hd}%
  \BibitemOpen
  \bibfield  {author} {\bibinfo {author} {\bibfnamefont {K.}~\bibnamefont
  {Choi}}, \bibinfo {author} {\bibfnamefont {K.~S.}\ \bibnamefont {Jeong}},
  \bibinfo {author} {\bibfnamefont {T.}~\bibnamefont {Kobayashi}}, \ and\
  \bibinfo {author} {\bibfnamefont {K.-i.}\ \bibnamefont {Okumura}},\ }\href
  {\doibase 10.1016/j.physletb.2005.11.078} {\bibfield  {journal} {\bibinfo
  {journal} {Phys. Lett. B}\ }\textbf {\bibinfo {volume} {633}},\ \bibinfo
  {pages} {355} (\bibinfo {year} {2006})},\ \Eprint
  {http://arxiv.org/abs/hep-ph/0508029} {arXiv:hep-ph/0508029} \BibitemShut
  {NoStop}%
\bibitem [{\citenamefont {Choi}\ \emph {et~al.}(2007)\citenamefont {Choi},
  \citenamefont {Jeong}, \citenamefont {Kobayashi},\ and\ \citenamefont
  {Okumura}}]{Choi:2006xb}%
  \BibitemOpen
  \bibfield  {author} {\bibinfo {author} {\bibfnamefont {K.}~\bibnamefont
  {Choi}}, \bibinfo {author} {\bibfnamefont {K.~S.}\ \bibnamefont {Jeong}},
  \bibinfo {author} {\bibfnamefont {T.}~\bibnamefont {Kobayashi}}, \ and\
  \bibinfo {author} {\bibfnamefont {K.-i.}\ \bibnamefont {Okumura}},\ }\href
  {\doibase 10.1103/PhysRevD.75.095012} {\bibfield  {journal} {\bibinfo
  {journal} {Phys. Rev. D}\ }\textbf {\bibinfo {volume} {75}},\ \bibinfo
  {pages} {095012} (\bibinfo {year} {2007})},\ \Eprint
  {http://arxiv.org/abs/hep-ph/0612258} {arXiv:hep-ph/0612258} \BibitemShut
  {NoStop}%
\bibitem [{\citenamefont {Chikashige}\ \emph {et~al.}(1981)\citenamefont
  {Chikashige}, \citenamefont {Mohapatra},\ and\ \citenamefont
  {Peccei}}]{Chikashige:1980ui}%
  \BibitemOpen
  \bibfield  {author} {\bibinfo {author} {\bibfnamefont {Y.}~\bibnamefont
  {Chikashige}}, \bibinfo {author} {\bibfnamefont {R.~N.}\ \bibnamefont
  {Mohapatra}}, \ and\ \bibinfo {author} {\bibfnamefont {R.~D.}\ \bibnamefont
  {Peccei}},\ }\href {\doibase 10.1016/0370-2693(81)90011-3} {\bibfield
  {journal} {\bibinfo  {journal} {Phys. Lett. B}\ }\textbf {\bibinfo {volume}
  {98}},\ \bibinfo {pages} {265} (\bibinfo {year} {1981})}\BibitemShut
  {NoStop}%
\bibitem [{\citenamefont {Gelmini}\ and\ \citenamefont
  {Roncadelli}(1981)}]{Gelmini:1980re}%
  \BibitemOpen
  \bibfield  {author} {\bibinfo {author} {\bibfnamefont {G.~B.}\ \bibnamefont
  {Gelmini}}\ and\ \bibinfo {author} {\bibfnamefont {M.}~\bibnamefont
  {Roncadelli}},\ }\href {\doibase 10.1016/0370-2693(81)90559-1} {\bibfield
  {journal} {\bibinfo  {journal} {Phys. Lett. B}\ }\textbf {\bibinfo {volume}
  {99}},\ \bibinfo {pages} {411} (\bibinfo {year} {1981})}\BibitemShut
  {NoStop}%
\bibitem [{\citenamefont {Li}\ and\ \citenamefont {Yu}(2023)}]{Li:2023kuz}%
  \BibitemOpen
  \bibfield  {author} {\bibinfo {author} {\bibfnamefont {S.-P.}\ \bibnamefont
  {Li}}\ and\ \bibinfo {author} {\bibfnamefont {B.}~\bibnamefont {Yu}},\
  }\href@noop {} {\  (\bibinfo {year} {2023})},\ \Eprint
  {http://arxiv.org/abs/2310.13492} {arXiv:2310.13492 [hep-ph]} \BibitemShut
  {NoStop}%
\bibitem [{\citenamefont {Escudero}\ and\ \citenamefont
  {Witte}(2021)}]{Escudero:2021rfi}%
  \BibitemOpen
  \bibfield  {author} {\bibinfo {author} {\bibfnamefont {M.}~\bibnamefont
  {Escudero}}\ and\ \bibinfo {author} {\bibfnamefont {S.~J.}\ \bibnamefont
  {Witte}},\ }\href {\doibase 10.1140/epjc/s10052-021-09276-5} {\bibfield
  {journal} {\bibinfo  {journal} {Eur. Phys. J. C}\ }\textbf {\bibinfo {volume}
  {81}},\ \bibinfo {pages} {515} (\bibinfo {year} {2021})},\ \Eprint
  {http://arxiv.org/abs/2103.03249} {arXiv:2103.03249 [hep-ph]} \BibitemShut
  {NoStop}%
\bibitem [{\citenamefont {Akita}\ and\ \citenamefont
  {Niibo}(2023)}]{Akita:2023qiz}%
  \BibitemOpen
  \bibfield  {author} {\bibinfo {author} {\bibfnamefont {K.}~\bibnamefont
  {Akita}}\ and\ \bibinfo {author} {\bibfnamefont {M.}~\bibnamefont {Niibo}},\
  }\href {\doibase 10.1007/JHEP07(2023)132} {\bibfield  {journal} {\bibinfo
  {journal} {JHEP}\ }\textbf {\bibinfo {volume} {07}},\ \bibinfo {pages} {132}
  (\bibinfo {year} {2023})},\ \Eprint {http://arxiv.org/abs/2304.04430}
  {arXiv:2304.04430 [hep-ph]} \BibitemShut {NoStop}%
\bibitem [{\citenamefont {Ade}\ \emph {et~al.}(2019)\citenamefont {Ade} \emph
  {et~al.}}]{SimonsObservatory:2018koc}%
  \BibitemOpen
  \bibfield  {author} {\bibinfo {author} {\bibfnamefont {P.}~\bibnamefont
  {Ade}} \emph {et~al.} (\bibinfo {collaboration} {Simons Observatory}),\
  }\href {\doibase 10.1088/1475-7516/2019/02/056} {\bibfield  {journal}
  {\bibinfo  {journal} {JCAP}\ }\textbf {\bibinfo {volume} {02}},\ \bibinfo
  {pages} {056} (\bibinfo {year} {2019})},\ \Eprint
  {http://arxiv.org/abs/1808.07445} {arXiv:1808.07445 [astro-ph.CO]}
  \BibitemShut {NoStop}%
\bibitem [{\citenamefont {Abazajian}\ \emph {et~al.}(2016)\citenamefont
  {Abazajian} \emph {et~al.}}]{CMB-S4:2016ple}%
  \BibitemOpen
  \bibfield  {author} {\bibinfo {author} {\bibfnamefont {K.~N.}\ \bibnamefont
  {Abazajian}} \emph {et~al.} (\bibinfo {collaboration} {CMB-S4}),\ }\href@noop
  {} {\  (\bibinfo {year} {2016})},\ \Eprint {http://arxiv.org/abs/1610.02743}
  {arXiv:1610.02743 [astro-ph.CO]} \BibitemShut {NoStop}%
\bibitem [{\citenamefont {Akrami}\ \emph {et~al.}(2020)\citenamefont {Akrami}
  \emph {et~al.}}]{Planck:2018jri}%
  \BibitemOpen
  \bibfield  {author} {\bibinfo {author} {\bibfnamefont {Y.}~\bibnamefont
  {Akrami}} \emph {et~al.} (\bibinfo {collaboration} {Planck}),\ }\href
  {\doibase 10.1051/0004-6361/201833887} {\bibfield  {journal} {\bibinfo
  {journal} {Astron. Astrophys.}\ }\textbf {\bibinfo {volume} {641}},\ \bibinfo
  {pages} {A10} (\bibinfo {year} {2020})},\ \Eprint
  {http://arxiv.org/abs/1807.06211} {arXiv:1807.06211 [astro-ph.CO]}
  \BibitemShut {NoStop}%
\bibitem [{\citenamefont {Harigaya}\ \emph {et~al.}(2015)\citenamefont
  {Harigaya}, \citenamefont {Ibe}, \citenamefont {Kawasaki},\ and\
  \citenamefont {Yanagida}}]{Harigaya:2015hha}%
  \BibitemOpen
  \bibfield  {author} {\bibinfo {author} {\bibfnamefont {K.}~\bibnamefont
  {Harigaya}}, \bibinfo {author} {\bibfnamefont {M.}~\bibnamefont {Ibe}},
  \bibinfo {author} {\bibfnamefont {M.}~\bibnamefont {Kawasaki}}, \ and\
  \bibinfo {author} {\bibfnamefont {T.~T.}\ \bibnamefont {Yanagida}},\ }\href
  {\doibase 10.1088/1475-7516/2015/11/003} {\bibfield  {journal} {\bibinfo
  {journal} {JCAP}\ }\textbf {\bibinfo {volume} {11}},\ \bibinfo {pages} {003}
  (\bibinfo {year} {2015})},\ \Eprint {http://arxiv.org/abs/1507.00119}
  {arXiv:1507.00119 [hep-ph]} \BibitemShut {NoStop}%
\bibitem [{\citenamefont {Adame}\ \emph {et~al.}(2024)\citenamefont {Adame}
  \emph {et~al.}}]{DESI:2024mwx}%
  \BibitemOpen
  \bibfield  {author} {\bibinfo {author} {\bibfnamefont {A.~G.}\ \bibnamefont
  {Adame}} \emph {et~al.} (\bibinfo {collaboration} {DESI}),\ }\href@noop {} {\
   (\bibinfo {year} {2024})},\ \Eprint {http://arxiv.org/abs/2404.03002}
  {arXiv:2404.03002 [astro-ph.CO]} \BibitemShut {NoStop}%
\bibitem [{\citenamefont {Workman}\ \emph {et~al.}(2022)\citenamefont {Workman}
  \emph {et~al.}}]{ParticleDataGroup:2022pth}%
  \BibitemOpen
  \bibfield  {author} {\bibinfo {author} {\bibfnamefont {R.~L.}\ \bibnamefont
  {Workman}} \emph {et~al.} (\bibinfo {collaboration} {Particle Data Group}),\
  }\href {\doibase 10.1093/ptep/ptac097} {\bibfield  {journal} {\bibinfo
  {journal} {PTEP}\ }\textbf {\bibinfo {volume} {2022}},\ \bibinfo {pages}
  {083C01} (\bibinfo {year} {2022})}\BibitemShut {NoStop}%
\bibitem [{\citenamefont {Allanach}\ and\ \citenamefont
  {Haber}(2024)}]{Allanach:2024suz}%
  \BibitemOpen
  \bibfield  {author} {\bibinfo {author} {\bibfnamefont {B.}~\bibnamefont
  {Allanach}}\ and\ \bibinfo {author} {\bibfnamefont {H.~E.}\ \bibnamefont
  {Haber}},\ }\href@noop {} {\  (\bibinfo {year} {2024})},\ \Eprint
  {http://arxiv.org/abs/2401.03827} {arXiv:2401.03827 [hep-ph]} \BibitemShut
  {NoStop}%
\end{thebibliography}%
\bibliographystyle{apsrev4-1}
	
\clearpage
\newpage

\end{document}